# Expected Utility Networks

**Piero La Mura**
Graduate School of Business
Stanford University
Stanford, CA 94305, USA
plamura@stanford.edu

**Yoav Shoham**
Dept. of Computer Science
Stanford University
Stanford, CA 94305, USA
shoham@cs.stanford.edu

## Abstract

We introduce a new class of graphical representations, expected utility networks (EUNs), and discuss some of its properties and potential applications to artificial intelligence and economic theory.

In EUNs not only probabilities, but also utilities enjoy a modular representation. EUNs are undirected graphs with two types of arc, representing probability and utility dependencies respectively. The representation of utilities is based on a novel notion of conditional utility independence, which we introduce and discuss in the context of other existing proposals.

Just as probabilistic inference involves the computation of conditional probabilities, *strategic inference* involves the computation of conditional expected utilities for alternative plans of action. We define a new notion of conditional expected utility (EU) independence, and show that in EUNs node separation with respect to the probability and utility subgraphs implies conditional EU independence.

## 1 Introduction

Modularity is the cornerstone of knowledge representation in AI; it allows concise representations of otherwise quite complex concepts. Logic offers modularity via the compositional nature of the logical connectives, and the property is exploited by theorem provers. Probability allows this via the notion of probabilistic independence, a notion fully exploited by Bayesian networks. In recent years there have been several attempts to provide modular utility representations of preferences (Bacchus and Grove 1995, Doyle and Wellman 1995, Shoham 1997).

It has proven difficult to devise a useful representation of utilities; this difficulty can certainly be ascribed to the different properties of utility and probability functions, but also, more fundamentally, to the fact that reasoning about probabilities and utilities together requires more than simply gluing together a representation of utility and one of probability.

In fact, just as probabilistic inference involves the computation of conditional probabilities, *strategic inference* – the reasoning process which underlies rational decision-making[1] – involves the computation of conditional expected utilities for alternative plans of action, which may not have a modular representation even if probabilities and utilities, taken separately, do.

The purpose of this paper is to introduce a new class of graphical representations, expected utility networks (EUNs). EUNs are undirected graphs with two types of arc, representing probability and utility dependencies respectively. In EUNs not only probabilities, but also utilities enjoy a modular representation. The representation of utilities is based on a novel notion of conditional utility independence, which departs significantly from other existing proposals, and is defined in close analogy with its probabilistic counterpart.

We also define a novel notion of conditional expected utility (EU) independence, and show that in EUNs node separation with respect to the probability and utility subgraphs implies conditional EU independence. In this respect, choosing the "right" notion of conditional utility independence turns out to be crucial.

What's important about conditionally independent decisions is that they can be effectively decentralized: a single, complicated agent can be replaced by simpler, conditionally independent sub-agents, who can do just as well. This property is of interest not only to artifi-cial intelligence, since it can be exploited to reduce the complexity of planning, but also to economic theory, as it suggests a principled way for the identification of optimal task allocations within economic organizations.

---

[1] Here, and elsewhere, the term "strategic" is used in the context of individual decision-making, and does not necessarily refer to a multi-agent scenario.

The rest of the paper is organized as follows. In section 2 we introduce our notion of conditional utility independence, and discuss it in the context of other recent proposals in the literature. In section 3 we formally introduce EUNs, and discuss some of their structural properties. Next, we extend the utility function from elementary "states" or outcomes to general events, with the interpretation that the utility of an event is the expected utility of that event, conditional on it being true. We explain this, and other characteristics of the underlying decision theoretic framework, in section 4. In section 5 we show that conditional probability and utility independence jointly imply conditional expected utility independence, and argue that conditionally independent decisions can be effectively decentralized. In section 6 we address the issue of probabilistic and strategic inference in EUNs, and show how conditional probabilities and conditional expected utilities can be recovered from the structural elements of EUNs. In the last section, we discuss a concrete application of the EUN methodology in the context of an economic example: a second-price auction.

## 2 Conditional independence of probabilities and utilities

Conditional probability independence is a powerful notion: it incorporates a natural, intuitive notion of relevance, and may dramatically reduce the complexity of probabilistic inference by allowing a convenient decomposition of the probability function.

In strategic inference, reducing the complexity of the decision problem calls for a decomposition of utilities along with probabilities. Yet, this is generally not enough: even if probabilities and utilities are separately decomposable, strategic inference typically involves computation of the *expected utilities* for alternative plans of action, and hence what is really important is the ability to decompose the expected utility function.

Several proposals which recently appeared in the literature (Bacchus and Grove 1995, Shoham 1997) rely on *additive* notions of utility independence, while the familiar notion of probabilistic independence is *multiplicative*. This difference may account for the difficulty encountered by these proposals to achieve a convenient decomposition of the expected utility function, and hence an effective reduction in the complexity of strategic inference.

In this section, we propose a multiplicative notion of conditional utility independence, which is a close analogue of its probabilistic counterpart. In the following sections, we argue that these two notions turn out to play well together, by inducing a modular decomposition of the expected utility function, and a consequent simplification of the decision process.

Let $\{X_i\}_{i \in N}$ ($N = \{1, ..., n\}$) be a finite, ordered set of random variables[2], and let $x^0 = (x_1^0, ..., x_n^0)$ be some arbitrary given realization which will act as the reference point (we use uppercase letters to denote random variables, and lowercase to denote their realizations). A joint realization $x = (x_1, ..., x_n)$ represents a (global) *state*, or *outcome*. For any $M \subset N$, we denote by $X_M$ the set $\{X_i\}_{i \in M}$. Let $p$ be a strictly positive probability measure defined on the Boolean algebra $\mathcal{A}$ generated by $X_N$, and let $u$ be a (utility) function which assigns to each state $x$ a positive real number. We assume that the decision maker's beliefs and preferences are completely characterized by $(p, u)$. Specifically, we assume that $p$ represents the decision maker's prior beliefs, and that for any two probability measures $p'$ and $p''$, $p' \succ p''$ ($p'$ is preferred to $p''$) if and only if $E_{p'}(u) > E_{p''}(u)$, where $E_p(u) = \sum_x u(x)p(x)$. Finally, let

$$q(x_M|x_{N-M}) = \frac{p(x_M, x_{N-M})}{p(x_M^0, x_{N-M})}.$$

The interpretation of $q$ is in terms of *ceteris paribus* comparisons: it tells us how the probability changes when the values of $X_M$ are shifted away from the reference point, while the values of $X_{N-M}$ are held fixed at $x_{N-M}$.

We also define a corresponding *ceteris paribus* comparison operator for utilities:

$$w(x_M|x_{N-M}) = \frac{u(x_M, x_{N-M})}{u(x_M^0, x_{N-M})}$$

One way to interpret $w$ is as a measure of the intensity of preference for $x_M$ (with respect to the reference point) conditional on $x_{N-M}$.

Suppose that $q(x_M|x_{N-M})$ only depends on $x_K$, where $K \subset N - M$, but not on $x_{N-M-K}$. It is easily verified that this condition holds for all $x_N$ if and only if $X_M$ is probabilistically independent of $X_{N-M-K}$ given $X_K$. We express (and record) this by defining new quantities $q(x_M|x_K) = q(x_M|x_{N-M})$, where the conditions $x_{N-M-K}$ are dropped. We call this notion *p-independence*: note that it is only defined in terms of states (as opposed to general events), and corresponds to conditional probability independence whenever it is defined. Specifically, if $A, B$ and $C$ are three subsets of the set $X_N$ of all random variables, the statement "$A$ is *p*-independent of $B$ given $C$" only makes sense if $A, B$ and $C$ constitute a partition of $X_N$, and in that case it is equivalent to the statement that $A$ and $B$ are probabilistically independent given $C$.

---

[2]To keep the notation simple, we assume that they may take only finitely many values. Yet, the construction is easily extended to more general classes of random variables.

A corresponding notion of conditional utility independence (*u-independence*) is defined accordingly. Suppose that $w(x_M|x_{N-M})$ depends on $x_K$, but not on $x_{N-M-K}$. Hence, the intensity of preference for the variables in $X_M$ (relative to their reference values) depends on the values of $X_K$, but not on those of $X_{N-M-K}$. In that case, we again define new quantities $w(x_M|x_K) = w(x_M|x_{N-M})$, and say that $X_M$ is *u*-independent of $X_{N-M-K}$ given $X_K$.

It is instructive to compare our notion of conditional utility independence with several other proposals which have appeared in the literature, in the context of an example adapted from Bacchus and Grove (1995). Suppose that there are two basic events, $H$ and $W$ ("health and wealth"), and that the following payoff tables, where payoffs are expressed as multiples of $u(\neg H \cap \neg W)$ (an arbitrary reference point), represent the decision maker's preferences over $H$ and $W$ in two different scenarios:

|      | $\neg W$ | $W$ |
| ---- | -------- | --- |
| $\neg H$ | 1    | 2   |
| $H$  | 3        | 6   |

|      | $\neg W$ | $W$ |
| ---- | -------- | --- |
| $\neg H$ | 1    | 2   |
| $H$  | 3        | 4   |

Bacchus and Grove's utility independence is, fundamentally, a qualitative notion, which in the example reduces to payoff dominance. Since $H$ dominates $\neg H$ and $W$ dominates $\neg W$, utility independence holds in both cases.

Additive utility independence specializes utility independence by requiring that probability measures with the same marginals be indifferent. In our 2 × 2 example, this amounts to the restriction that $u(H \cap W) + u(\neg H \cap \neg W) = u(H \cap \neg W) + u(\neg H \cap W)$. Hence, additive utility independence holds in the second case but not in the first.

Shoham further specializes the notion of additive independence, with the following intended interpretation: the two Boolean variables in the example are independent if it is possible to associate to each of them a linear "contribution", such that the utility of a joint realization is given by the sum of the contributions. In our 2 × 2 example, this criterion coincides with additive independence; however, in the general case it is more stringent.

We too introduce quantitative information, but in a different way: the two variables in the example are independent in our sense (conditional on the empty set) if the increment in utility relative to the reference point is the product of the increments along each component. The intended interpretation of utility independence in our case is that the "intensity of preference" for one variable with respect to its reference value, represented by the *ceteris paribus* utility ratio, does not depend on the particular value taken by the other variable. Hence, in our sense, $H$ and $W$ are independent in the first scenario but not in the second.

We claim that *u*-independence is a particularly attractive notion for two reasons:

- it is information that is natural to elicit from people, as it purely involves relevance considerations and order-of-magnitude comparisons between utilities

- it gives rise to a graphical representation and associated inference mechanism, expected utility networks (defined in the next section), which is simultaneously modular in probabilities, utilities and expected utilities.

## 3  Expected utility networks: a formal definition and some structural properties

We define an expected utility network as an undirected graph $G$ with two types of arc, representing probability and utility dependencies respectively. Each node represents a random variable (say, $X_i$), and is associated with two positive functions, $q(x_i|x_{P(i)})$ and $w(x_i|x_{U(i)})$, where $P(i)$ denotes the set of nodes directly connected to $X_i$ via probability arcs[3], and $U(i)$ the corresponding set of nodes directly connected to $X_i$ via utility arcs.

These quantities are interpreted as the probability and utility ratios (defined in the previous section) produced by some expected utility representation $(p, u)$, and may be assessed by the decision maker through *ceteris paribus* comparisons. Alternatively, the probability layer of a EUN may be initially specified as a Bayes network, and the probability ratios $q$ derived from conditional probability tables.

Figure 1 depicts a simple EUN. Although it is possible to present much richer examples, we select this one

---

[3]$P(i)$ corresponds to the *Markov mantle* of $X_i$, i.e., the minimal set of variables such that, conditional on those, $X_i$ is (probability) independent of everything else.

because it ties in with an auction example discussed in the last section.

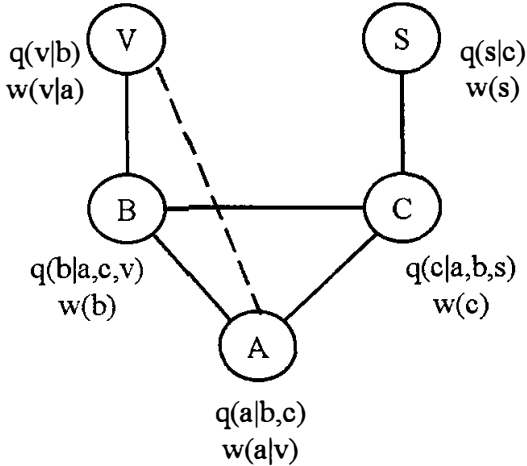

Figure 1: A simple EUN. Probability arcs are represented by solid lines, and utility arcs by dashed lines.

If the $q$ and $w$ functions are specified directly, then any arbitrary assignment of positive functions $q(x_i | x_{BP(i)}, x^0_{AP(i)})$ for all $i$ (where $AP(i)$ denotes the set of all variables in $P(i)$ whose index is greater than $i$, and $BP(i) = P(i) - AP(i)$) uniquely identifies a corresponding probability function. Similarly, any arbitrary assignment of positive functions $w(x_i | x_{BP(i)}, x^0_{AP(i)})$ identifies a utility function, unique up to normalization.[4]

We remark that, if $q(x_i|x_{-i})$ only depends on $x_{P(i)}$, then fixing $x_{P(i)}$ completely specifies the behavior of the probability function along the $i - th$ coordinate (up to the probability of the reference point), and that such behavior does not depend on the particular values taken by the other variables. The same is true about the utility of $x_i$ with respect to its reference value, given $x_{U(i)}$.

It turns out that node separation with respect to the probability and utility subgraphs characterizes all the implied p- and u- independencies. More precisely, for any probability – utility pair $(p, u)$ there exists an undirected graph $G$ such that, if $A$, $B$ and $C$ are three subsets of variables (each variable being associated with a node in the graph), $A$ is p- (resp., u-) independent of $B$ given $C$ if and only if $C$ separates $A$ from $B$ with respect to the probability (resp., utility) subgraph (in the sense that every path from $A$ to $B$ in the subgraph must pass through $C$). In the language of Pearl (1988), $G$ is a perfect map of the independence structure.

---

[4]The particular normalization we adopt is discussed in section 4.

**Theorem 1** *The set of p- and u- independencies generated by any pair $(p, u)$ has a perfect map.*

**Proof.** We follow the methodology of Bacchus and Grove (1995), that is, we appeal to a necessary and sufficient condition in Pearl and Paz (1989) and check that suitable generalizations of p-independence and u-independence both possess the following five properties: symmetry, decomposition, intersection, strong union and transitivity.

We prove it in the case of utility; the proof for probability is analogous. Let $A, B, C, D, R, R', R''$ be subsets of random variables, where $R$, $R'$ and $R''$ always denote the subset of remaining variables in the appropriate context (so, for instance, in the context of some $A, B$ and $C$, $R = X_N - A - B - C$). As elsewhere in this paper, we use uppercase/lowercase to denote (subsets of) random variables and their realizations respectively.

For the purpose of this proof, let's say that $A$ is independent of $B$ given $C$, and write $I(A, B|C)$ if and only if $w(a|b, c, r) = w(a|b_0, c, r)$ for all $(a, b, c, r)$. Then the following properties hold.

**Symmetry**: $I(A, B|C) \Rightarrow I(B, A|C)$.

This follows because

$$w(b|a,c,r) = \frac{u(a,b,c,r)}{u(a,b^0,c,r)} = \frac{u(a,b,c,r)}{u(a^0,b,c,r)} \frac{u(a^0,b,c,r)}{u(a,b^0,c,r)} = \frac{u(a,b^0,c,r)}{u(a^0,b^0,c,r)} \frac{u(a^0,b,c,r)}{u(a,b^0,c,r)} = w(b|a^0,c,r).$$

**Decomposition**: $I(A, B \cup D|C) \Rightarrow I(A, B|C) \wedge I(A, D|C)$.

This is equivalent to saying that $w(a|b,c,d,r) = w(a|b_0, c, d_0, r)$ implies $w(a|b, c, r') = w(a|b_0, c, r')$ and $w(a|c, d, r'') = w(a|c, d_0, r'')$. This follows trivially, because $r' = (d, r)$ and $r'' = (b, r)$.

**Intersection**: $I(A, B|C \cup D) \wedge I(A, D|B \cup C) \Rightarrow I(A, B \cup D|C)$.

Equivalently, $w(a|b, c, d, r) = w(a|b_0, c, d, r)$ and $w(a|b, c, d, r) = w(a|b, c, d_0, r)$ imply $w(a|b, c, d, r) = w(a|b_0, c, d_0, r)$. This also follows quite easily by algebraic manipulation.

**Strong union**: $I(A, B|C) \Rightarrow I(B, A|C \cup D)$.

Equivalently, $w(a|c, b, r) = w(a|b_0, c, r)$ implies $w(b|a, c, d, r') = w(b|a_0, c, d, r')$. This follows by symmetry, and the fact that $r = (d, r')$.

**Transitivity**: $I(A, B|C) \Rightarrow I(A, V|C) \vee I(B, V|C)$, where $V$ is any single variable. This is equivalent to saying that $w(a|b, c, r) = w(a|b_0, c, r)$ implies $w(a|v, c, r') = w(a|v_0, c, r')$ or $w(b|v, c, r'') = w(b|v_0, c, r'')$, which follows by observing that either $v$ is in $b$ or else is in $r$. ∎

## 4   Conditional expected utility

While probability is a set function, defined for general events, utility is so far only defined for elementary events (states). The notion of $p$-independence introduced in section 2 precisely corresponds to conditional probability independence whenever it is defined, and it is only defined in terms of states – in turn, this enabled us to define a corresponding notion of conditional utility independence, also defined in terms of states, which we named $u$-independence.

As we have seen, all $p$- and $u$- independencies can be immediately recovered from the graphical structure of EUNs, because they are fully characterized by node separation with respect to the probability and utility subgraphs. For instance, in the simple EUN represented in figure 1, $V$ and $C$ are conditionally $p$- and $u$- independent of each other, given $A, B$ and $S$.

Suppose now that $A$ and $B$ are controllable variables, in the sense that their values can be freely set by the decision maker. A rational decision maker will want to choose values of $A$ and $B$ which maximize expected utility; hence, for each assignment $(a, b)$, the decision maker should compute the corresponding expected utility, and identify an optimal decision. Clearly, the decision process becomes quite cumbersome when there are many decision variables; to reduce its complexity, we seek conditions under which the expected utility calculations can be conveniently decomposed.

The first step will be to extend utility to a be a set function as well, with the following interpretation: the utility of an event is the expected utility of the event, conditional on the event being true. Formally,

$$u(E) = \sum_{x \in E} u(x) p(x|E).$$

The following important property is an immediate consequence of the definition: for any nonempty $E \in \mathcal{A}$ and for any non-empty, finite partition $\{E_k\}$ of $E$, where the $E_k$ may or may not be elementary "states",

$$u(E) = \sum_k u(E_k) p(E_k|E).$$

The (von Neumann - Morgenstern) utilities we start from[5] are only defined up to positive affine transformations. It is natural then to normalize the utility measure around certain values, just as probabilities are normalized to lie between zero and one. Hence, we require that $u(True) = 1$, where $True$ denotes the tautological event, or the entire universe.[6]

Although it won't play a direct role in EUNs, in order to facilitate the exposition we also define the *value* (or *impact*) of event $E$:

$$v(E) = u(E) p(E)$$

Under the above normalization $v$ is a (strictly positive) probability measure, since it is an additive set function, and

$$v(True) = 1 \geq v(E) > 0$$

for all nonempty $E$. Moreover, since $p$ is also strictly positive, we have that

$$u(E) = \frac{v(E)}{p(E)}.$$

Note the remarkable structure of conditional expected utility: the utility "measure" is simply the ratio of two probability measures, one representing value, and the other belief.

Beside being important for the practical construction of EUNs, this normalization of $u$ allows us to speak about "good" and "bad" events. $True$ – the status quo – is neutral, neither good or bad. An event $E$ is said to be good (i.e., better than $True$) if $u(E) > 1$, and bad if $u(E) < 1$.

The conditional versions of the three set functions – probability, utility, and value – are defined in the natural way: $p(E|F) = p(E \cap F)/p(F)$, and similarly $u(E|F) = u(E \cap F)/u(F)$, and $v(E|F) = v(E \cap F)/v(F)$.

The three notions of conditioning are related by

$$v(E|F) = u(E|F) p(E|F).$$

Being a probability measure, $p$ obeys Bayes' rule (and, clearly, so does $v$):

$$p(F|E) = \frac{p(E|F) p(F)}{p(E|F) p(F) + p(E|\text{-}F) p(\text{-}F)}$$

---

[5] We start with von Neumann - Morgenstern utilities and a given prior probability, as it is customary in economics. It is beyond the scope of this paper to discuss the decision-theoretic foundations of EUNs; we refer to La Mura and Shoham (1998) for a formal exposition of the underlying decision-theoretic framework, and a representation result for conditional (and counterfactual), hierarchical, multi-agent preferences.

[6] This normalization uniquely identifies the expected utility function if the utility of a second event $E_0$ is also fixed, or, equivalently, if utilities are expressed as multiples of $u(x^0)$, the utility of an arbitrary reference point, as we do in EUNs.

Bayes' rule does not hold for utilities, but a modified version of it does:

$$u(F|E) = \frac{u(E|F)u(F)}{u(E|F)u(F)p(F|E) + u(E|\neg F)u(\neg F)p(\neg F|E)}$$

Note that this is a "hybrid" relationship: conditional utility depends, among other things, on conditional probabilities. This is another fact which is important to keep in mind in connection with EUNs.

## 5 Conditional expected utility independence

We have now extended the utility function from complete states to arbitrary events, but this new concept will be useful only insofar as it can be associated with a corresponding independence notion, and this extended notion is also captured in the structure of the graph. In this section we show that both these conditions hold. First we shall define a notion of conditional expected utility independence, and then show that this notion is indeed captured in the graphical structure of EUNs.

We define *conditional expected utility independence* (or, more concisely, conditional EU independence) for general events in analogy with the familiar notion of conditional probability independence. Two events, $E$ and $F$, are said to be conditionally EU independent given a third event $G$ if

$$u(E \cap F | G) = u(E|G)u(F|G).$$

Conditional expected utility independence generalizes $u$-independence from states to general events, much as conditional probability independence generalizes $p$-independence. Yet, since expected utilities involve probabilities as well, the relationship between conditional EU independence and $u$-independence is mediated by probabilities.

Let's look at the general case first. Consider a partition of the set of all random variables into three subsets $A$, $B$ and $C$. The conditional expected utility of $b$ given $a$ is

$$u(b|a) = \frac{u(a,b)}{u(a)} = \frac{\sum_c u(a,b,c)p(c|a,b)}{\sum_{b,c} u(a,b,c)p(b,c|a)}.$$

Suppose now that $a$ separates $b$ from $c$ with respect to both the probability and utility subgraphs. Then the following simplification obtains:

$$u(b|a) = \frac{w(b|a)}{\sum_b w(b|a)p(b|a)}$$

$$p(b|a) = \frac{q(b|a)}{\sum_b q(b|a)}.$$

Hence, the formula for $u(b|a)$ does not involve terms in $C$, and similarly $u(c|a)$ does not involve terms in $B$.

This is not true if $B$ and $C$ are not $p$-independent, as the following example shows.

**Example 1** *Consider the special case in which $A$ is empty, and $w(b) = 1$ (in which case, we say that $B$ is payoff-irrelevant). Then $B$ and $C$ are $u$-independent, although they may not be $p$-independent.*

*Hence, $\frac{u(b)}{u(b')} = \frac{\sum_c w(c)q(c|b)}{\sum_c w(c)q(c|b')}$, a quantity which is generally different from one. Intuitively, in this case $B$ is purely instrumental to $C$: it is irrelevant in itself, but its expected utility reflects the influence that a particular choice of $B$ has on the probability of $C$. If $B$ and $C$ are also $p$-independent, then the above expression reduces to $\frac{u(b)}{u(b')} = 1$ (in which case, we say that $B$ is strategically irrelevant).*

These observations are central to the following result.

**Theorem 2** *$p$- and $u$- independence jointly imply conditional expected utility independence.*

Hence, the graphical structure of EUNs can be exploited to identify conditional EU independencies: node separation with respect to both the utility and probability subgraph implies conditional EU independence. The upshot is that, conditional on $A$, decisions regarding $B$ and $C$ can be effectively decomposed: if both $B$ and $C$ contain variables which are under the control of the decision maker, it is not necessary nor useful to possess information about $C$ in order to decide on $B$, and vice versa. One way to think about such decomposability is in terms of strategic decentralization: a single, centralized decision maker can be replaced by two conditionally independent, simpler agents, who only need to worry about their own respective domains in order to make jointly optimal decisions.

## 6 Inference in expected utility networks

In EUNs, probabilities and utilities are implicitly described by the $q$ and $w$ functions, together with the topological structure of the network. Probabilistic inference involves the computation of conditional probabilities, and strategic inference the computation of conditional expected utilities; in this section, we show how these quantities can be readily recovered from the structural elements of EUNs.

The probability layer of a EUN is essentially a Markov network, even though probabilities are subject to a somewhat unusual normalization, and hence probabilistic inference can be performed with standard techniques whenever the potentials $q$ are known. In turn, potentials can either be assigned directly by the decision maker in the form of *ceteris paribus* comparisons, or derived from conditional probabilities if one starts with a Bayes network.

The advantage of using utility "potentials" in EUNs is that they are based purely on utility comparisons between states, which do not involve probabilities: this enables one to elicit all the relevant preferences from the decision maker without assuming that he or she already knows the probabilities.

Although we don't tackle here the issue of computational efficiency for probabilistic and strategic inference in EUNs (a topic which is next on our research agenda), we'll show how the two fundamental operations of marginalization and conditionalization for probabilities and expected utilities can be easily reduced to operations on the probability and utility potentials.

Once the potentials are known, the computation of $p(x)/p(x^0)$ is straightforward:

$$\frac{p(x)}{p(x^0)} = \frac{p(x_1, x_{-1}^0)}{p(x^0)} \frac{p(x)}{p(x_1, x_{-1}^0)}$$

$$= q(x_1|x_{P(1)}^0) \frac{p(x_2, x_1, x_{-\{1,2\}}^0)}{p(x_2^0, x_1, x_{-\{1,2\}}^0)} \frac{p(x)}{p(x_2, x_1, x_{-\{1,2\}}^0)}$$

$$= q(x_1|x_{P(1)}^0) q(x_2|x_1, x_{AP(2)}^0) \frac{p(x)}{p(x_2, x_1, x_{-\{1,2\}}^0)}$$

$$= \ldots = \times_i q(x_i|x_{BP(i)}, x_{AP(i)}^0).$$

One can obtain $p(x_M)/p(x^0)$ (where $p(x_M)$ is now the marginal probability function for a subset of random variables $X_M$) by summing over the $X_{N-M}$:

$$\frac{p(x_M)}{p(x^0)} = \sum_{x_{N-M}} \frac{p(x_M, x_{N-M})}{p(x^0)}$$

One can then use $p(x_M)/p(x^0)$ to compute ratios of marginal probabilities $p(x_A)/p(x_B)$, and in particular conditional probabilities $p(x_A|x_B) = p(x_A, x_B)/p(x_B)$.

To compute $u(x)/u(x^0)$, we use the same decomposition:

$$\frac{u(x)}{u(x^0)} = \times_i w(x_i|x_{BU(i)}, x_{AU(i)}^0)$$

The marginal (expected) utility of $x_M$, relative to the reference point, can hence be computed as

$$\frac{u(x_M)}{u(x^0)} = \sum_{x_{N-M}} \frac{u(x_M, x_{N-M})}{u(x^0)} p(x_{N-M}|x_M)$$

One can then use ratios $u(x_M)/u(x^0)$ to compute ratios of expected utilities $u(x_A)/u(x_B)$, and in particular conditional expected utilities $u(x_A|x_B) = u(x_A, x_B)/u(x_B)$.

Notice that marginal utilities generally depend on probabilities. For instance, the utility of catching a particular cab rather than not is measured by the ratio $u(Cab)/u(\neg Cab)$, and will generally depend on how likely it is that another cab will show up, on the probability of rain, and so on.

Again, we remark that using utility "potentials" as the initial data in EUNs (rather than conditional expected utilities) enables the decision maker to specify all the relevant preferences without any prior knowledge of the probabilities.

## 7  Example: second-price auction

To conclude our presentation of EUNs, we propose a concrete application of the EUN machinery in the context of an economic example: a second-price ("Vickrey") auction.

In a second-price auction the highest bidder gets the auctioned good, but only pays the second-highest bid. We assume that there are two bidders, and that the values of the good to each bidder correspond to the realizations of independent random variables.

Agent 1 privately observes her own value for the good (denoted by $V$), and then decides how much to bid ($B$). Independently, the value of the good for agent 2 ($S$) is realized, and contingent on that he decides how much to bid ($C$). The two bids jointly determine the final allocation ($A$), which is a pair $a = (g, m)$ denoting who gets the good ($g = 1, 2$) and how much must be paid for it ($m$).

To remove potential confusion (as our presentation of EUNs was centered on a single-agent perspective), we emphasize that we only model the game from the point of view of a single agent, and solve it as an individual decision problem. Yet, in a second price auction, as well as in other dominance-solvable games, this also suffices to identify the unique equilibrium.

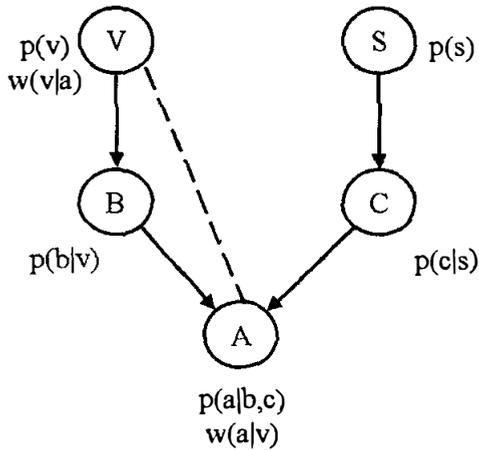

Figure 2: An independent-value, second-price auction, from the point of view of agent 1.

Figure 2 represents the auction from the point of view of agent 1. The probability layer is represented as a Bayes network, to emphasize the causal structure of the events in the game. Once again, we remark that the probability potentials $q$ (and the corresponding Markov representation) can be readily derived from the conditional probability tables, in which case the resulting EUN looks like the one depicted in Figure 1. Here we omit this extra step, as we won't need it in the context of the example (since we shall simplify the expected utility function directly, rather than appealing to Theorem 2). Also, we omit all the utility potentials $w$ which are identically equal to 1 (corresponding to payoff-irrelevant variables).

The probability of ending up with a particular allocation $(g, m)$ depends on $b$ and $c$, and is given by[7]

$$p(a|b,c) = \begin{cases} 1 \text{ if } b \geq c, c = m, g = 1 \\ 1 \text{ if } b < c, b = m, g = 2 \\ 0 \text{ otherwise} \end{cases}$$

For the purpose of exposition, we also postulate a particular functional form for agent 1's preferences. We assume that the following condition holds:

$$w(a|v) = \begin{cases} \frac{1+v}{1+m} \text{ if } g = 1 \\ 1 \text{ otherwise} \end{cases}$$

Note that agent 1's preferences on different allocations depend on her realized value for the good. We also assume that the distribution of agent 2's bids, from the point of view of agent 1, has full support.

---
[7]For definiteness, we assume that in the case of identical bids agent 1 gets the good.

Agent 1 chooses her bid in order to maximize utility, given her private value for the good. The expected utility of $b$ conditional on $v$ is given by:

$$\begin{aligned} u(b|v) &= \int u(a,b,c,s|v) p(da, dc, ds|b, v) \\ &= \frac{u(x^0)}{u(v)} \int w(a,b,c,s,v) p(da|b,c) p(dc, ds) \\ &= \frac{u(x^0) w(v|a^0)}{u(v)} \left( \int_{-\infty}^{b} \frac{1+v}{1+c} p(dc) + \int_{b}^{\infty} p(dc) \right). \end{aligned}$$

The first-order condition for optimality is given by

$$\left. \frac{1+v}{1+c} p(c) \right|_{c=b^*} = \left. p(c) \right|_{c=b^*}$$

and returns $b^* = v$.

Hence, regardless of what her opponent is going to bid (as long as the distribution has full support), the optimal strategy for agent 1 is to bid her true evaluation: i.e., to bid exactly the amount of money which keeps her indifferent between getting the good (and paying for it) or not.

### References

F. Bacchus and A. Grove (1995), Graphical models for preference and utility. In *Proc. 11th Conference on Uncertainty in Artificial Intelligence*, pp. 3-10.

J. Doyle and M. P. Wellman (1995), Defining preferences as Ceteris Paribus Comparatives. In *Proc. AAAI Spring Symp. on Qualitative Decision Making*, pp. 69-75.

P. La Mura and Y. Shoham (1998), Conditional, hierarchical, multi-agent preferences. In *Proc. of Theoretical Aspects of Rationality and Knowledge – VII*, pp. 215-224.

J. Pearl (1988), *Probabilistic reasoning in intelligent systems*. Morgan Kaufmann.

J. Pearl and A. Paz (1989), Graphoids: A Graph-Based Logic for Reasoning About Relevance Relations. In B. Du Boulay (Ed.), *Advances in Artificial Intelligence - II*, North-Holland.

Y. Shoham (1997), A Symmetric View of Probabilities and Utilities. In *Proc. 13th Conference on Uncertainty in Artificial Intelligence*, pp. 429-436.